\documentclass[12pt]{article}
\usepackage[pdftex]{graphicx}

\begin{document}
\author{G.G.Kozlov}
\title{
Spectral Dependence of Degree of Localization of Eigenfunctions of the 1D Schrodinger Equation with a
Peacewise-Constant Random Potential}
\maketitle
\begin{abstract}
The perturbation theory is developed for joint statistics of the advanced and
 retarded Green's functions of the 1D Schrodinger equation with a
 piecewise-constant random potential. Using this method, analytical
 expressions are obtained for spectral dependence of the degree of
 localization and for the limiting (at $t\rightarrow\infty$) probability to find
 the particle at the point it was located at $t = 0$ (Andeson criterion).
 Definition of the localization length is introduced. The computer
 experiments confirming correctness of the calculations are described.
\end{abstract}

\section{Introduction, Formulation of the Problem, and Main Results}

Mathematical problems arising in the physics of solid-state random systems are characterized by a
complexity and absence of universal methods of analysis. In searching for such methods, an important
role is played by strongly simplified models of disordered systems, with the 1D single-particle ones
being the most important among them. The heuristic significance of the 1D models is, however, not the
only one. The physical systems like J-aggregates, quantum wells, optical fibers, Bragg layered
structures, etc. can be directly described by the 1D models. In such systems, one may expect strong
effects of disorder, which makes studying of the disordered 1D models especially topical.

        In the theory of the simplest solid-state disordered models, one can distinguish
         the {\it continuous} and
         {\it discrete} models\cite{LGP}. The continuous models employ the Schrodinger
        equation   $[-d^2/dx^2+{\cal U}(x)]\psi=E\psi$
   with one or another potential ${\cal U}(x)$, while the discrete models use
        random matrices of a model Hamiltonian. In spite of similarity of these two models,
        each of them has its own specificity. For example, when analyzing vibrations
        of disordered chains \cite{Belous1,Belous2}, the discrete model is used, whereas when studying
        propagation of electromagnetic waves in disordered layered structures, the
        continuous model looks more convincing.

For the 1D models, in a number of cases, mathematically correct methods of theoretical analysis can be
proposed \cite{Dyson,Ber1,LGP}. Of particular interest are the cases, when one and
 the same method appears to be
suitable for analysis of several different model problems, and the method proves to be, to a certain
extent, universal. It is noteworthy, in this connection, that the perturbation theory for the joint
statistics of the advanced and retarded Green's functions used in this paper for analysis of the
continuous disordered model, has been successfully used previously for studying the discrete models
\cite{Koz1,Koz2}.

Let us pass to the problem
 studied in this paper. Consider a 1D continuous disordered model with
a peacewise-constant random potential ${\cal U}(x)$ equal to $u+\varepsilon_n$
 inside the intervals $x\in[b(n-1),bn], n=1,2,...,N$. Here, $\varepsilon_n$ are the independent
limited random quantities with a known distribution function $P(\varepsilon)$, and $u<0$
 is the negative number
sufficiently big to make ${\cal U}(x)<0$ at $x\in[0,Nb]$.
 The length $b$ is a specified parameter of the potential ${\cal U}(x)$.
  For $x\bar\in[0,Nb]$, we assume that ${\cal U}(x)=0$.
   The distribution function $P(\varepsilon)$ is taken in the following, fairly general, form
  \begin{equation}
  P(\varepsilon)={1\over\Delta}p\bigg({\varepsilon\over\Delta}\bigg),\hskip3mm
  p(\varepsilon)>0,\hskip3mm M_n\equiv\int p(\varepsilon)\varepsilon^nd\varepsilon,\hskip3mm M_0=1,\hskip3mm M_1=0
  \label{22}
  \end{equation}
The parameter $\Delta$ is the measure of disorder and, at $\Delta = 0$, the function  ${\cal U}(x)$
 represents a
potential box with a flat bottom with the depth $u$ and length $Nb$. At  $\Delta>0$, we can say that the
function ${\cal U}(x)$ corresponds to a potential box with a fluctuating bottom. Hereafter, we imply the
tyhermodynamic limit  $N\rightarrow\infty$.

        Consider the motion of a particle in such a random potential and formulate
        the following problem. Let the particle, at $t = 0$, to be located in the
        point $r=Nb$, (i.e., at the right side of the potential box with the
        fluctuating bottom) , and we are seeking for the density of probability
        that the particle will remain in this point at $t\rightarrow\infty$. From the mathematical
         viewpoint, it means that, at $t = 0$, the wave function of the particle had the
          form $\Psi(t=0,x)=\delta(x-r)$, and we have to find
          $D=\lim_{t\rightarrow\infty}\langle|\Psi(t,r)|^2\rangle$.
           The  angle brackets  here and below indicate
          averaging over realizations of the random potential ${\cal U}(x)$. This problem is well
          known in the theory of Anderson localization \cite{LGP,And} and it can be shown $\cite{Koz1}$ that,
          if $\psi_n(x)$ are the eigenfunctions of the Hamiltonian   $H=-d^2/dx^2+{\cal U}(x)$,
           then $D=\langle\sum_n|\psi_n(r)|^4\rangle$. Nonzero value of $D$
          indicates that there exist localized functions (i.e., functions essentially nonzero
           in some finite region, with its size independent of $N$, at $N\rightarrow\infty$) among
            eigenfunctions of the Hamiltonian  $H$ \cite{Koz1,LGP,And}. To judge about the presence
            or absence of the localized states in the energy interval $[U,U+dU]$, in \cite{Koz1}
            there has been introduced the participation function $W(U)$ defined by
            the relationship $W(U)dU=\langle\sum_{n, E_n\in[U,U+dU]}|\psi_n(r)|^4\rangle$,
             where $E_n$ is the eigenenergy of the Hamiltonian $H$.
            If the eigenfunctions  of the Hamiltonian $H$ with energy $U$
            are delocalized, then $W(U) = 0$.
            Otherwise, $W(U)$ is nonzero. For this reason, the participation function
            will be below referred to as {\it spectral dependence of the degree of
            localization.} The main results of this paper are the following expressions
            for the function $W$ and the quantity $D$:
\begin{equation}
W(U)=\Theta(-U)\Theta(U-u)\hskip2mm\bigg({\Delta\over u}\bigg)^2{M_2\over 2\pi}
\hskip2mm
{\sin^2b\sqrt{U-u}\over b\sqrt{U-u}}+O(\Delta^3),
\hskip5mm u<0
\label{91}
\end{equation}
\begin{equation}
D=\int_u^0 W(U)dU=
\bigg({\Delta\over u}\bigg)^2{M_2\over 2\pi b}\bigg(\sqrt{-u}-{\sin[2b\sqrt{-u}]\over 2b}\bigg)+O(\Delta^3),
\label{92}
\end{equation}
These formulas are applicable to the above 1D continuous model with a peacewise-constant random potential.

        Concluding the introduction, note that the Helmholtz equation describing
        propagation of electromagnetic waves in a layered system, in fact, coincides
        with the Schrodinger equation studied in this paper. This gives the grounds
        to assert that the results obtained in this paper can be used in studies of
        propagation of electromagnetic waves in 1D photonic crystals in the presence
        of disorder.

\section{Continuous model. General properties of the Schredinger equation Green's function.}

To solve the above typical problem of the theory of disordered systems, we will apply the method of
joint statistics of the advanced and retarded edge Green's function (EGF) used in \cite{Koz1,Koz2}
 for analysis of
the discrete 1D disordered models. A crucial point of the above method is the fact that, in the
discrete 1D model, the EGF of a chain with a single structural unit added can be expressed
algebraically through the EGF of the initial chain \cite{Dyson,LGP,Koz2}.
  In this section, we will briefly remind the
properties of Green's function of the differential Schrodinger equation with the Hamiltonian
\begin{equation}
H\equiv-d^2/dx^2+{\cal U}(x)
\label{1}
\end{equation}
 and, in the next one, we will present a similar relation
 for the EGF, valid for the case of the continuous
model with the peacewise-constant potential ${\cal U}(x)$.

       Green's function of operator (\ref{1}) is defined by the formula
 \begin{equation}
 G_{xx'}(\Omega)\equiv \sum_n{\psi_n(x)\psi_n^\ast (x')\over\Omega-E_n}+
 \int dp {\phi_p(x)\phi^\ast _{p}(x')\over \Omega-{\cal E}_p}
 \label{5}
 \end{equation}

where $\psi_n (\phi_p)$ and $E_n ({\cal E}_p)$ are,
 respectively, the eigenfunctions and eigenvalues of  operator (\ref{1}),
corresponding to the discrete (continuous) spectrum. Here, $n (p)$ is the discrete (continuous) number of
the eigenfunction. The energy argument $\Omega$ of the Green's function is, generally, a complex number
$\Omega=U-\imath V$ (with $U$ and $V$ being real). Using Eq. (\ref{5}),
 one can show that the solution $\Psi(t,x)$ of the time-dependent Schrodinger equation,
with the initial condition $\Psi(0,x)=\Psi_0(x)$, can be expressed in terms of
 the Green's function (\ref{5}), as
follows
\begin{equation}
\Psi(t,x)=\lim_{V\rightarrow+0}\hskip3mm{1\over 2\pi\imath}\int dU dx' e^{\imath U t}\hskip2mm
 G_{xx'}(U-\imath V)\hskip2mm\Psi_0(x'),\hskip5mm t>0
\label{7}
\end{equation}
 It can be easily shown that the Green's function (\ref{5}) satisfies the differential
equation
\begin{equation}
 \bigg[\Omega+{d^2\over dx^2}-{\cal U}(x)\bigg]\hskip1mm G_{xx'}(\Omega)=\delta(x-x')
 \label{8}
 \end{equation}
 where only solutions {\it vanishing} at $x\rightarrow\pm\infty$ should be taken to provide convergence of
integrals (\ref{7}). Taking into accunt the above properties of the Green's function, one can see that the
quantity $D$, introduced in Sect.1, is expressed through the product of diagonal elements of the
advanced and retarded EGF in the following way
\begin{equation}
D=\lim_{t\rightarrow\infty}\langle|\Psi(t,r)|^2\rangle=\lim_{t\rightarrow\infty}
\lim_{V_{1,2}\rightarrow +0}\hskip2mm{1\over 4\pi^2}\int dU_1dU_2\hskip1mm e^{\imath(U_1-U_2)t}
\langle G_{rr}(U_1-\imath V_1)G_{rr}(U_2+\imath V_2)\rangle,
 \label{17}
 \end{equation}
 Hear $r$ is the coordinate of the right side of the
potential box with a fluctuating bottom described in Introduction. Using  spectral expansion (\ref{5}),
one can show that the quantity $D$ is determined only by discrete states of the Hamiltonian (\ref{1}),
 with the
following formula being valid \cite{Koz1}
 \begin{equation}
D=\lim_{t\rightarrow\infty}\hskip2mm\langle|\Psi(t,r)|^2\rangle=
\bigg\langle\sum_{n}|\psi_n(r)|^4\bigg\rangle
\label{19}
 \end{equation}
 By limiting  the region of integration in (\ref{17}) so that $U_{1,2}\in [U,U+dU]$,
   we can  obtain the participation function $W(U)$, introduced in \cite{Koz1}
   \begin{equation}
  W(U)dU=\bigg\langle\sum_{E_n\in[U,U+dU]}|\psi_n(r)|^4\bigg\rangle
\label{20}
  \end{equation}
     As was already mentioned, nonzero value of $W(U)$ indicates presence of
     localized states in the energy interval $[U,U+dU]$.
In the opposite case, $W(U)=0$.

\section{Case of the peacewise-constant potential. Recurrent relations for the EGF.}

Consider the following family of random
  potentials constant within the intervals of length $b$:
  \begin{equation}
{\cal U}_m(x)=\cases{u+\varepsilon_n, \hbox{ если } x\in[b(n-1),bn],
\hbox{ при } x\le mb\hskip5mm (u<0, n \hbox{ -- целое })\cr
0, \hbox{ если } x>mb}
\label{21}
\end{equation}

Here $\varepsilon_n$  are the independent bounded random
quantities with the distribution function $P(\varepsilon)$ (\ref{22}), and $u$ is the
negative number sufficiently large to meet the condition $u+\varepsilon_n<0$.
 Assume the EGF $G_{mb,mb}^m(\Omega)\equiv \gamma_m(\Omega)$ of the Schrodinger
equation (\ref{1}) with the potential ${\cal U}_m(x)$ to be known.
Let us pass to the potential ${\cal U}_{m+1}(x)$ and consider
the EGF $G_{b(m+1),b(m+1)}^{m+1}(\Omega)\equiv\gamma_{m+1}(\Omega)$ of Eq. (4)
 corresponding to this potential ${\cal U}_{m+1}(x)$. In this section, we will express the EGF
 $\gamma_{m+1}(\Omega)$ through  $\gamma_{m}(\Omega)$
 using the fact that, at $x<mb$, these potentials are the same.

 Note, first of all, that the
discrete spectrum of operator (\ref{1}) with potential (\ref{21}),
 we are interested in, is positioned on the
negative semiaxis. For this reason, in what follows, we will consider real part $U$ of the energy argument
of the Green's function to be negative $\Omega=U-\imath V, U<0$.
 The Green's function $G^m_{x,bm}(\Omega)$ meets the equations
 \begin{equation}
\cases{[\Omega+d^2/dx^2-{\cal U}_m(x)]G^m_{x,bm}(\Omega)=0, \hbox{ at } x<bm\cr
[\Omega+d^2/dx^2]G^m_{x,bm}(\Omega)=0, \hbox{ at } x>bm}
\label{23}
\end{equation}
Let us introduce the functions $\Psi_{\pm}(x)$, so that
\begin{equation}
[\Omega+d^2/dx^2-{\cal U}_m(x)]\Psi_-(x)=0, \hbox{ with }\Psi_-(-\infty)=0 \hbox{ и } \Psi_-(mb)=1
\label{24}
\end{equation}
\begin{equation}
[\Omega+d^2/dx^2]\Psi_+(x)=0, \hbox{ with }\Psi_+(\infty)=0 \hbox{ и } \Psi_+(bm)=1
\label{25}
\end{equation}

 Equations and conditions (\ref{24}) and (\ref{25}) determine the functions $\Psi_\pm(x)$
   in a unique way. It
follows from Eq. (\ref{25}) that
 \begin{equation}
\Psi_+(x)=e^{\imath\sqrt{\Omega}[x-bm]},  \hskip10mm \hbox{  } \Omega=U-\imath V,  \hskip2mm V>0,  \hskip2mm U<0
\label{26}
\end{equation}

  The Green's function we are interested in can be expressed in terms of
the functions $\Psi_\pm(x)$ as follows
 \begin{equation}
G^m_{x,bm}(\Omega)=\cases{A\Psi_-(x), \hbox{ at } x<mb\cr B\Psi_+(x), \hbox{ at } x>mb}
\label{28}
\end{equation}
  The  continuity of the function  $G^m_{x,bm}(\Omega)$ at $x=mb$  together with  a
unit jump of its derivative in this point lead to the system of equations for the constants $A$ and $B$.
Solving this system and taking into consideration that $\Psi_{\pm}(mb)=1$,
 we obtained for the EGF $G^m_{bm,bm}(\Omega)$ the
following relation
\begin{equation}
\gamma_m(\Omega)\equiv G^m_{bm,bm}(\Omega)={1\over\Psi'_+(bm)-\Psi'_-(bm)}
\label{31}
\end{equation}
 Note that the function $\Psi_+(x)$, entering this relation, is known in the explicit
form (\ref{26}).

Now, let us pass from the potential ${\cal U}_m(x)$ to the potential ${\cal U}_{m+1}(x)$
 and consider the Green's
function $G_{x,b(m+1)}^{m+1}(\Omega)$. It satisfies the equations similar to (\ref{23})
\begin{equation}
\cases{[\Omega+d^2/dx^2-{\cal U}_m(x)]G^{m+1}_{x,b(m+1)}(\Omega)=0, \hbox{ at } x<bm\cr
[\Omega+d^2/dx^2-\eta]G^{m+1}_{x,b(m+1)}(\Omega)=0, \hbox{ at } bm<x<b(m+1),\hskip5mm\eta\equiv
u+\varepsilon_{m+1}\cr [\Omega+d^2/dx^2]G^{m+1}_{x,b(m+1)}(\Omega)=0, \hbox{ at } x>b(m+1)}
\label{32}
\end{equation}
  By analogy with (\ref{28}), we can write the following expressions for $G_{x,b(m+1)}^{m+1}(\Omega)$
\begin{equation}
\cases{G^{m+1}_{x,b(m+1)}(\Omega)=\tilde A\Psi_-(x), \hbox{ at } x<bm\cr
G^{m+1}_{x,b(m+1)}(\Omega)=C e^{\imath\sqrt{\Omega-\eta}x}+F e^{-\imath\sqrt{\Omega-\eta}x}, \hbox{ at
} bm<x<b(m+1)\cr G^{m+1}_{x,b(m+1)}(\Omega)=\tilde B\Psi_+(x), \hbox{ at } x>b(m+1)}
\label{33}
\end{equation}

At $x = mb$, the Green's function $G_{x,b(m+1)}^{m+1}(\Omega)$
 should be continuous together with its first derivative,
while, at $x=(m+1)b$, the function $G_{x,b(m+1)}^{m+1}(\Omega)$
 should be continuous, and its derivative should experience a unit
jump.
This yields four equations for the constants $\tilde A, C,F$, and $\tilde B$ entering (\ref{33}).
 The EGF $\gamma_{m+1}(\Omega)$ of
interest, corresponding to the potential ${\cal U}_{m+1}(x)$,
can be obtained from (\ref{33})
\begin{equation}
 \gamma_{m+1}(\Omega)\equiv G^{m+1}_{b(m+1),b(m+1)}(\Omega)=\tilde B\Psi_+(b(m+1))
 \label{39}
 \end{equation}

  Finding the constant $\tilde B$
from the above system of equations for $\tilde A,C,F$, and $\tilde B$ and using Eq. (\ref{26})
 for the function $\Psi_+(x)$, we
obtain for the EGF $\gamma_{m+1}(\Omega)$ the following relation
\begin{equation}
{t\hskip1mm\sqrt{\Omega-\eta} -\Psi_-'(mb) \over
\sqrt{\Omega-\eta} +t\hskip1mm\Psi_-'(mb) }=
-\imath\sqrt{\Omega\over\Omega-\eta}+
{1\over\gamma_{m+1}(\Omega)\sqrt{\Omega-\eta}}\hskip4mm\hbox{ где
} t\equiv \hbox{ tg }[b\sqrt{\Omega-\eta}] \label{40}
\end{equation}

 Now, using (\ref{31}) we can express $\Psi_-'(mb)$ through the EGF
$\gamma_m(\Omega)$
 \begin{equation}
\Psi_-'(mb)=\imath\sqrt{\Omega}-{1\over\gamma_m(\Omega)}
\label{41}
\end{equation}

  Here, we took into consideration that $\Psi'_+(bm)=\imath\sqrt\Omega$ at $U<0$.
   With the use of Eqs. (\ref{40}) and (\ref{41}), we
can obtain the sought relation between EGF $\gamma_{m+1}(\Omega)$ and EGF $\gamma_m(\Omega)$
 (the corresponding operation will be further
referred to as ${\cal R}^{-1}$)
\begin{equation}
\gamma_{m+1}={h+\gamma_m\over q+v\gamma_m}
\equiv {\cal R}_{\Omega,\eta}^{-1}(\gamma_m)
\label{44}
\end{equation}
where
$$
h\equiv-{t\over \sqrt{\Omega-\eta}+\imath t\sqrt\Omega},\hskip5mm q\equiv{\sqrt{\Omega-\eta}-\imath
t\sqrt\Omega\over\sqrt{\Omega-\eta}+\imath t\sqrt\Omega}\hskip5mm v\equiv-{t\eta\over
\sqrt{\Omega-\eta}+\imath t\sqrt\Omega}
$$
$$
t=\hbox{tg}\bigg[b\sqrt{\Omega-\eta}\bigg], \hskip5mm \Omega=U\pm\imath V,\hskip5mm U<0, V=+0,
\hskip5mm\eta\equiv u+\varepsilon_{m+1}
$$

 Below, we will need the operation ${\cal R}$  inverse to (\ref{44}), which has the
following form
\begin{equation}
\gamma_m={\gamma_{m+1} q-h\over 1-v\gamma_{m+1}}\equiv {\cal R}_{\Omega,\eta}(\gamma_{m+1})
\label{45}
\end{equation}
 Thus, the whole method of analysis of the joint statistics of the EGF, developed
in \cite{Koz1,Koz2} for discrete models, can be applied to the considered case of a continuous model,
corresponding to Schrodinger equation (\ref{1}) with the peacewise-constant potential (\ref{21}).
 Relevant calculations
 are presented in the following sections.

\section{Calculating spectral dependence of the degree of localization}

\subsection{Joint statistics of the Green's functions}

Spectral dependence of the degree of localization $W(U)$ and the probability $D$ to find the particle at
the edge of the random 1D system under consideration are given, respectively, by Eqs. (\ref{20}) and (\ref{19}).
These formulas are identical to those  for similar quantities of the discrete model\cite{Koz1,Koz2}. For
this reason, for realisation of Eqs. (\ref{20}) and (\ref{19}),
 one can use the method based on calculation of
joint statistics of the advanced and retarded Green's functions developed in \cite{Koz1,Koz2}. Let us remind
briefly this method. In the integrals entering Eq. (\ref{17}), we make the following substitution
$\omega\equiv U_2-U_1, U\equiv U_1$ and
change the notations $G_{rr}\rightarrow \gamma$. Then, for the value $D$ (\ref{19}),
 we can write the following expression
 \begin{equation}
D={1\over 4 \pi^2}\lim_{t\rightarrow\infty}\lim_{V_{1,2}\rightarrow+0}\int d\omega dU e^{\imath\omega
t}\langle\gamma(U-\imath V_1)\gamma(U+\omega+\imath V_2)\rangle
\label{63}
\end{equation}
  If the function $\rho(x_1y_1x_2y_2)$
 is the joint statistics of the advanced and retarded Green's functions entering (\ref{63}) (here,
the arguments $x_i$ and $y_i$, $i=1,2$ correspond to real and imagenary parts of these functions),
 then the averaged
value of their product can be represented in the form
 \begin{equation}
\langle\gamma(U-\imath V_1)\gamma(U+\omega+\imath
V_2)\rangle=\int dx_1dy_1dx_2dy_2\rho(x_1y_1x_2y_2)[x_1x_2-y_1y_2+\imath(x_1y_2+x_2y_1)]\equiv
\label{64}
\end{equation}
$$
\equiv \langle x_1x_2\rangle-\langle y_1y_2\rangle+\imath\langle x_1y_2\rangle+\imath\langle
y_1x_2\rangle,
$$

  It was shown in \cite{Koz1} that it suffices to
calculate only $\imath\langle x_1y_2\rangle$  and to multiply the result by 4.
In accordance with \cite{Koz2}, calculation of this
contribution at $V_{1,2}\rightarrow+0$
 can be performed using the formula
   \begin{equation}
  \langle x_1y_2\rangle=\int d\varepsilon d\tilde x_1 d\tilde x_2
  P(\varepsilon)\sigma_{U,U+\omega}(\tilde x_1\tilde x_2) x_1(\tilde x_1)
  y_2(\tilde x_2),
\label{65}
  \end{equation}
 with the form of the dependences $ x_1(\tilde x_1)$ and $y_2(\tilde x_2)$
 being determined by the fractional-linear function ${\cal R}_{\Omega,\eta}^{-1}(x)$ (\ref{44})
 \footnote {We will present this function using notations of \cite{Koz2}}:
\begin{equation}
{\cal R}_{\Omega,\eta}^{-1}(x)={h+x\over q+vx}
\equiv
{a_{\Omega,\eta}+b_{\Omega,\eta} x\over c_{\Omega,\eta}+g_{\Omega,\eta} x},
\hskip5mm\hbox{ }
\label{66}
\end{equation}
which looks like this
\begin{equation}
x_1=\hbox{Re }\bigg[{\cal R}_{\Omega_1,\eta}^{-1}(\tilde x_1)\bigg],\hskip1mm y_2=\hbox{Im }\bigg[{\cal
R}^{-1}_{\Omega_2,\eta}(\tilde x_2)\bigg],\hskip1mm
\eta=u+\varepsilon, \hskip1mm
\Omega_1=U-\imath V_1, \hskip1mm \Omega_2=U+\omega+\imath V_2
\label{67}
\end{equation}
 Function $\sigma_{U,U+\omega}(x_1x_2)$  in Eq. (\ref{65})
  represents the joint statistics of the {\it real} Green's functions with the
energy arguments $U$ and $U+\omega$, respectively. In \cite{Koz2},
 it has been shown that this function satisfies the
following equation
 \begin{equation}
\sigma_{U_1U_2}(x_1x_2)=\int d\varepsilon  \hskip1mm P(\varepsilon)
\hskip1mm\sigma_{U_1U_2}\bigg[{\cal R}_{U_1\eta}(x_1),{\cal R}_{U_2\eta}(x_2)\bigg]
\bigg|{d{\cal R}_{U_1\eta}(x_1)\over d
x_1}\bigg|\bigg|{d{\cal R}_{U_2\eta}(x_2)\over dx_2}\bigg|,\hskip1mm \eta=u+\varepsilon
\label{68}
\end{equation}
  with the operation ${\cal R}_{U,\eta}(x)$, in this case, being determined by Eq. (\ref{45}). As shown in
\cite{Koz2}, at $V_{1,2}\rightarrow+0$, the relations (\ref{67})
 lead to the following expressions for $y_2(\tilde x_2)$ and $x_1(\tilde x_1)$:
\begin{equation}
y_2(\tilde x_2)\bigg|_{V_2\rightarrow +0}=
\pi{a_{U+\omega,\eta} g_{U+\omega,\eta}-b_{U+\omega,\eta} c_{U+\omega,\eta}\over g_{U+\omega,\eta}^2}
\delta\bigg(\tilde x_2+{c_{U+\omega,\eta}\over g_{U+\omega,\eta}}\bigg),\hskip2mm
\label{69}
\end{equation}
\begin{equation}
x_1(\tilde x_1)\bigg|_{V_1\rightarrow +0}= {a_{U,\eta}+b_{U,\eta} \tilde x\over c_{U,\eta}+g_{U,\eta}
\tilde x}
\label{70}
\end{equation}

    Substitution
of these expressions into (\ref{65}), allows us to obtain, for the quantity $\langle x_1y_2\rangle$
 of interest the following
relationship \cite{Koz2}
  \begin{equation}
\langle x_1y_2\rangle=\pi\lim_{a\rightarrow\infty} a^2\int \sigma_{U U+\omega}(x,a) x dx
\label{71}
 \end{equation}
  When deriving this relationship, we took into account that the function $\sigma_{U,U+\omega}(x_1x_2)$
meets Eq. (\ref{68}).
It follows from the above formulas that the quantity $D$, we are interested in, may be
represented in the form
\begin{equation}
D={\imath\over \pi^2}\lim_{V_{1,2}\rightarrow 0, t\rightarrow\infty}
\int e^{\imath\omega t}\langle x_1y_2\rangle d\omega dU=
{\imath\over \pi}\lim_{a\rightarrow \infty, t\rightarrow\infty}
\int e^{\imath\omega t}a^2\sigma_{U,U+\omega}(x,a)x dx d\omega dU
\label{72}
\end{equation}
 As shown in \cite{Koz1}, the participation function $W(U)$ can be obtained from Eq.
(\ref{72}) by omitting the integration over $U$:
 \begin{equation}
W(U)={\imath\over \pi}\lim_{a\rightarrow \infty, t\rightarrow\infty}
\int e^{\imath\omega t}a^2\sigma_{U,U+\omega}(x,a)x dxd\omega
\label{73}
\end{equation}
  In this way, the problem is reduced to solving Eq. (\ref{68}).
The perturbative approach to equations of the type (\ref{68}), proposed in \cite{Koz1},
 is the power expansion in
$\Delta$ (see Eq. (\ref{22})), with the first nonzero correction being of the order of $\Delta^2$.
It was also shown
in \cite{Koz1} that, to calculate the quantities $D$ and $W(U)$,
 only the part of the solution of the equation for the
joint statistics (in our case, Eq. (\ref{68})), singular in $\omega$,
 is needed, with the singularity being of
the pole type.
Thus, the needed singular part (referred to as {\it sing}) can be represented in the form
 \begin{equation}
  \hbox{sing }\sigma_{U\omega}(x_1x_2)={\Delta^2\over \omega}{\cal F}_U(x_1x_2)+O(\Delta^3)
  \label{74}
  \end{equation}

Now, using Eq. (\ref{73}), for the function $W(U)$ and quantity $D$,
we obtain the following formulas
  \begin{equation}
 W(U)=-\Delta^2\lim_{a\rightarrow\infty}a^2\int{\cal F}_U(x,a)xdx +O(\Delta^3),
 \hskip10mm D=\int W(U)dU
\label{75}
  \end{equation}
 In the following section, we will present the perturbative approach to Eq. (\ref{68})
  and will derive an explicit
expression for the function ${\cal F}_U(x_1x_2)$ entering Eq. (\ref{74}).

\subsection{ Perturbative approach to Eq. (\ref{68})}

To solve the functional equations arising in the perturbation theory described below one has to find
the eigenfunctions and eigenvalues of the functional operator ${\cal H}_{\Omega,\eta}$,
 which acts upon  an
arbitrary function $f(x)$ as follows \footnote {This problem is solved in Appendix, and, in what follows,
we will use the results obtained in it.}
 \begin{equation}
{\cal H}_{\Omega,\eta}f(x)\equiv {d{\cal R}_{\Omega,\eta}\over dx} f[{\cal R}_{\Omega,\eta}(x)]
\label{46}
\end{equation}
where ${\cal R}_{\Omega,\eta}$  is given by (\ref{45}).
  We assume the parameter $\Delta$ to be small and represent
the sought function $\sigma_{U_1U_2}(x_1x_2)$ as a power series in $\Delta$.
 \begin{equation}
\sigma_{U_1U_2}(x_1x_2)=\sum_{n=0}^\infty Q_n(x_1,x_2)\Delta^n
\label{76}
\end{equation}
  Let us expand the function $\sigma_{U_1U_2}[{\cal R}_{U_1,\eta}(x_1),{\cal R}_{U_2,\eta}(x_2)] |{d{\cal R}_{U_1,\eta}(x_1)\over d
x_1}||{d{\cal R}_{U_2,\eta}(x_2)\over dx_2}|$, in the
right-hand side of (\ref{68}) into a power series in $\varepsilon$.
 Then, Eq. (\ref{68}) yields
   \begin{equation}
\sum_{n=0}^\infty Q_n(x_1,x_2)\Delta^n=
 \label{77}
 \end{equation}
 $$
 \sum_{n,l=0}^\infty{M_n\Delta^{n+l}\over n!}
{\partial^n\over\partial \varepsilon^n}\bigg\{Q_l
\bigg[{\cal R}_{U_1,\eta}(x_1),{\cal R}_{U_2,\eta}(x_2)\bigg]
\bigg|{d{\cal R}_{U_1,\eta}(x_1)\over d
x_1}\bigg|\bigg|{d{\cal R}_{U_2,\eta}(x_2)\over dx_2}\bigg|\bigg\}_{\varepsilon=0}
$$

   Remind that the
dependence on $\varepsilon$  in this equation is provided by the quantity $\eta=u+\varepsilon$.
 By equating the coefficients at
the same powers of $\Delta$ in the right- and left-hand sides of Eq. (\ref{77}), we obtain the recurrent
relations for the function $Q_n$
\begin{equation}
\Delta^0:\hskip10mm Q_0(x_1x_2)-Q_0
\bigg[{\cal R}_{U_1,u}(x_1),{\cal R}_{U_2,u}(x_2)\bigg]
\bigg|{d{\cal R}_{U_1,u}(x_1)\over
 dx_1}\bigg|\bigg|{d{\cal R}_{U_2,u}(x_2)\over dx_2}\bigg|=0
\label{78}
\end{equation}
 Since the first moment of the function $P(\varepsilon)$ (\ref{22}) is zero, the quantity $Q_1$
vanishes,
 \begin{equation}
 \Delta^2:\hskip10mm Q_2(x_1x_2)-Q_2
\bigg[{\cal R}_{U_1,u}(x_1),{\cal R}_{U_2,u}(x_2)\bigg]
\bigg|{d{\cal R}_{U_1,u}(x_1)\over
 dx_1}\bigg|\bigg|{d{\cal R}_{U_2,u}(x_2)\over dx_2}\bigg|=
\label{79}
 \end{equation}
$$
={M_2\over 2}{\partial^2\over\partial \varepsilon^2}\bigg\{Q_0
\bigg[{\cal R}_{U_1,\eta}(x_1),{\cal R}_{U_2,\eta}(x_2)\bigg]
\bigg|{d{\cal R}_{U_1,\eta}(x_1)\over d
x_1}\bigg|\bigg|{d{\cal R}_{U_2,\eta}(x_2)\over dx_2}\bigg|\bigg\}_{\varepsilon=0}\hskip5mm
\eta=u+\varepsilon
$$
  and so on.
From Eqs. (\ref{78}) and (\ref{79}), we see that they contain the
functional operator ${\cal H}_{U_i,u}, i=1,2$ (\ref{46}). Taking into account
its properties, described in Appendix, we can immediately write the solution of Eq. (\ref{78}) for $Q_0$:
\begin{equation}
 Q_0(x_1x_2)={\cal L}_{U_1,u}(x_1){\cal L}_{U_2,u}(x_2)
 \label{80}
 \end{equation}
To solve Eq. (\ref{79}), we will present the sought function $Q_2(x_1x_2)$
in the form of expansion over eigenfunctions
(\ref{62}) of operator (\ref{46}):
\begin{equation}
Q_2(x_1x_2)=\sum_{|n|+|l|\ne 0}C_{nl} s_n^{U_1,u}(x_1)s_l^{U_2,u}(x_2)
\label{81}
\end{equation}
  By substituting this series into the left-hand side of Eq. (\ref{79}) and by
expanding its right-hand side using (\ref{61}), we obtain, for the coefficients $C_{nl}$,
 the following formulas:
\begin{equation}
C_{nl}={1\over 1-\lambda_n(U_1,u)\lambda_l(U_2,u)}\hskip2mm {M_2\over
2}\hskip2mm{\partial^2\over\partial\varepsilon^2}\bigg[
J_n(U_1\hskip1mm\varepsilon)J_l(U_2\hskip1mm\varepsilon)
\bigg]_{ \varepsilon=0}
\label{82}
\end{equation}
where the quantities $J_n(U\hskip1mm\varepsilon)$ are defined as
\begin{equation}
J_n(U\hskip1mm\varepsilon)\equiv\int{{\cal L}_{U,u}({\cal R}_{U,\eta}(x))
\hskip1mm{\cal R}_{U,\eta}'(x)\over
{\cal G}_{U,u}^n(x)}\hskip1mm dx =\int{{\cal L}_{U,u}(z)
\over {\cal G}^n_{U,u}({\cal R}_{U,\eta}^{-1}(z))}\hskip1mm dz=J_{-n}^\ast(U\hskip1mm\varepsilon),
\hskip5mm\eta=u+\varepsilon
\label{83}
\end{equation}

   Definitions of the functions ${\cal L}$ and ${\cal G}$ entering these
expressions are given in Appendix. When expanding the right-hand side of (\ref{79}),
we used expression (\ref{80})
for the function $Q_0(x_1x_2)$.
 As was pointed out above, we are interested only in the part of $Q_2(x_1x_2)$ singular in $\omega=U_2-U_1$.
To extract  this part, one has to retain, in (\ref{81}), only the terms with $n = -l$ \cite{Koz1},
 since only for these
terms the denominator $ 1-\lambda_n(U_1,u)\lambda_l(U_2,u)$  in (\ref{82})
 turns into zero at $\omega=U_2-U_1=0$. The calculation identical to that
performed in \cite{Koz1} leads to the following expression for the function ${\cal F}_U(x_1x_2)$
 entering Eq. (\ref{75}):
\begin{equation}
{\cal F}_U(x_1x_2)=
-{\imath M_2\over 2b}\sqrt{U-u}
\sum_{n\ne 0}
{\partial^2\over\partial\varepsilon^2}\bigg| J_n(U\hskip1mm\varepsilon)
\bigg |_{ \varepsilon=0}^2
{s^{U,u}_n(x_1) s^{U,u}_{-n}(x_2)\over n}\hskip5mm
\label{84}
\end{equation}
  Now, let
us present explicit expressions for the integrals (\ref{83}):
\begin{equation}
J_0(U,\varepsilon)=1,\hskip1mm J_n(U\hskip1mm\varepsilon)
= {\cal G}^{-n}_{U,u}\bigg({\cal R}_{U,\eta}^{-1}(\bar r)\bigg)\bigg|_{\eta=u+\varepsilon}
=[J_1(U,\varepsilon)]^n,\hskip1mm n>0,\hskip1mm
\bar r={\sqrt{U}-\sqrt{U-u}\over\imath u}
\label{85}
\end{equation}

These expressions are obtained by
integrating (\ref{83}) with the help of residues. Note that, for calculations
of the derivatives entering Eq.
(\ref{82}), the value $\varepsilon$ can be considered so small that it does not
 affect positions of the poles of
the integrants with respect to the real axis (above or below).
Using Eq. (\ref{621}), we can obtain the relationship
\begin{equation}
{1\over {\cal G}_{U,u}^n({\cal R}_{U,u}^{-1}(z))}=\lambda_n(U){1\over {\cal
G}_{U,u}^n(z)}=\lambda_n(U,u)
\bigg({r-z\over r^\ast-z}\bigg)^n
\label{86}
\end{equation}
 which shows that $J_n(U,0)=0$  at $n\ne0$   and that,
 in the general case, the power expansion of ${\cal G}_{U,u}^{-1}({\cal R}^{-1}_{U,\eta}(\bar r))$
  in $\varepsilon$ starts from the first power and may be written in the form

\begin{equation}
{\cal G}_{U,u}^{-1}({\cal R}^{-1}_{U,\eta}(\bar r))\bigg|_{\eta=u+\varepsilon}=J_1(U,\varepsilon)
=K_U\varepsilon+O(\varepsilon^2)
\label{87}
\end{equation}

Substitution of this expression into (\ref{84}) shows that, in sum (\ref{84}),
only the terms with $J_{\pm 1}(U\varepsilon)$
survive, for which the second derivative of their module squared is nonzero at $\varepsilon = 0$.
 Thus, Eq.(\ref{84}) for the function  ${\cal F}_U(x_1x_2)$,
  can be represented in the form

\begin{equation}
{\cal F}_U(x_1x_2)=
-{\imath M_2\over b}\sqrt{U-u}
|K_U|^2\hskip2mm\bigg[ s^{U,u}_1(x_1) s^{U,u}_{-1}(x_2)
-s^{U,u}_{-1}(x_1) s^{U,u}_{1}(x_2)\bigg]
\label{88}
\end{equation}

  Direct algebraic calculations using
explicit expressions (\ref{61}) for function ${\cal G}_{U,u}(x)$ and (\ref{44})
 for the operation ${\cal R}^{-1}_{U,\eta}(x)$ show that
 \begin{equation}
K_U=\imath e^{\imath b\sqrt{U-u}}\hskip2mm {[\sqrt U-\sqrt{U-u}]^2\over 2 u(U-u)}\hskip1mm
\sin[b\sqrt{U-u}],\hskip10mm
|K_U|^2= {\sin^2[b\sqrt{U-u}]\over 4 (U-u)^2}
\label{89}
\end{equation}
  Finally, using
expressions for the moments and limiting values of the functions $\sigma^n_U(x)$,
 given in \cite{Koz1} (see Appendix),
with the aid of Eq. (\ref{60}),
 we can obtain the following expressions for the first moments and limiting
values of the $s$-functions:

\begin{equation}
\int s^{U,u}_n(x)xdx=\imath{n\over |n|}{\sqrt{U-u}\over u}\hskip2mm{|t|\over t}\hskip10mm
\lim_{a\rightarrow\infty}a^2 s^{U,u}_n(a)={1\over \pi}{\sqrt{U-u}\over u}{|t|\over t}
\label{90}
\end{equation}
 Then, using Eq. (\ref{75}), we obtain, for the participation function $W(U)$
and the quantity $D$,  expressions (\ref{91}) and (\ref{92}).

\section{Numerical experiment. Localization length.}

The most convincing way to verify theoretical results related to 1D solid-state disordered models is,
nowadays, to compare them with a numerical experiment. Below, we present the results of numerical
verification of Eq. (\ref{91}) for the participation function $W(U)$ (spectral dependence of the degree of
localization), which is considered to be the main result of this paper. In this verification, we used
definition (\ref{20}) at $dU << |u|$.
The wave functions entering Eq. (\ref{20}) were obtained by solving numerically
the edge problem for Schrodinger equation (\ref{1}) with the random potential (\ref{21})
 using the transfer matrix
technique. In the calculations, we assumed
$p(\varepsilon)=\Theta(\varepsilon+1/2)-\Theta(\varepsilon-1/2)$
 (see Eq. (\ref{22})) and the number of regions of
constant potential $N\sim 200 - 900$. The final function $W(U)$ was obtained by averaging over
$N_r\sim 2000 - 4000$
realizations of the random potential.
 When performing the above calculations, one should keep in mind the following: (i) As far as formula
(\ref{91}) obtained in this paper is valid in the thermodynamic limit, the number $N$
 should be sufficiently
large. However, at $N > 800 - 900$, in the calculations of the wave functions, the errors arising at
multiplications of a great number of the transfer matrices rapidly increase; (ii) For a given length
$bN$ of a random system, the degree of its disorder $\Delta$,
on the one hand, should be {\it large
enough} for the localization length to be smaller than $bN$, and, on the other, should be {\it small enough} not
to come out of the range of applicability of Eq. (\ref{91}); (iii) In these calculations, one has to check
quadratic character of the dependence of the computed function $W(U)$ and independence of the results on
$N$.

The results of numerical calculations for different values of the parameters $b$ and $\Delta$ are presented
in Fig.1 (the values of all the parameters are given in the figure), the smooth curves being
calculated using Eq.(\ref{91}) with no fitting. Figure 1b demonstrates better agreement between the theory
and experiment than Fig.1a, because, the above conditions were satisfied much better
 for the case of numerical dependence for Fig.1b.

In the numerical calculations, it is useful to be able to evaluate the localization length $l$ of the
wave functions for the random system with a given energy $U$. For such evaluations, one can use the
participation function $W(U)$ (\ref{91}) obtained in this paper. Consider the states
  of the random system with
the energies lying within the interval $[U,U+ dU]$.
The number of such states will be $\rho(U)dU$, where $\rho(U)$ is
the density of states. In virtue of spatial uniformity of statistical properties of the random
potential ${\cal U}(x)$, we can say that the "centers of gravity" of these localized states are distributed more
or less uniformly along the $x$ axis. Therefore, the number of states in the {\it energy} interval $[U,U+dU]$,
 whose
centers of gravity fall into the {\it spatial} interval $dL$ of the $x$-axis, can be estimated as
$[\rho(U)/L]dUdL$, where $L=Nb$
-- is the length of the potential box with a fluctuating bottom. Note that the participation function
(\ref{20}) is mainly contributed by the states whose centers of gravity are separated from the edge of the
potential box by the distance not exceeding their localization length $l$. The number $dn$
of such states is
estimated to be $dn=[\rho(U)/L]ldU$.
 By denoting the mean amplitude of these states at $x = 0$ as $\psi_U(0)$, we can write, for
the participation function (\ref{20}), the following approximate expression:

\begin{equation}
W(U)dU=|\psi_U(0)|^4dn\hskip10mm W(U)={\rho(U)l\over L}|\psi_U(0)|^4.
\label{93}
\end{equation}

 With the accuracy acceptable for
our purposes, we may assume that the density of states $\rho(U)$ entering this equation does not strongly
differ from that $\rho_0(U)$ for the potential box with the length $L$ and depth $u$ with {\it no
disorder}:
\begin{equation}
  \rho(U)\approx\rho_0(U)={L\over 2\pi \sqrt{U-u}}
  \label{94}
  \end{equation}
Then, formula (\ref{93}) yields
\begin{equation}
W(U)={l\hskip2mm |\psi_U(0)|^4\over 2\pi \sqrt{U-u}}
\label{95}
\end{equation}
The localization length $l$ entering this formula and the amplitude of the wave
function $\psi_U(0)$ can be connected by the normalization condition, which will provide a second relationship
for their calculation. If the quantity $\psi_U(0)$ were close to a {\it typical} amplitude of the wave function, the
above connection would have a simple form $|\psi_U(0)|^2l=1$.
 However, the arguments presented below show that the
amplitude of the wave function at $x = 0$ can be much smaller than its typical value, which we denote as
$\bar\psi_U$. Let us evaluate $\bar\psi_U$ based on the following reasoning.
 At $x < 0$, (i.e., outside the potential
box), the wave function has the form $\psi_U(x)=\psi_U(0)\exp[\sqrt{-U}x]$. At $0<x<l$,
scattering in the random potential is weak, and the
wave function, within this interval, approximately corresponds to free motion of the particle with the
energy $U$.
For this reason, for the wave function near the edge of the random system we can write the
following expressions
\begin{equation}
\cases{\psi_U(x)=\psi_U(0)\exp[x\sqrt{-U}]\hskip10mm x<0\cr
\psi_U(x)=A\sin[x\sqrt{U-u}+\varphi]\hskip10mm 0<x<l}
\label{96}
\end{equation}

 The energy $U$ is assumed here to be sufficiently high, so that the motion of
the particle has a ballistic, rather than tunnel, character $U-u-\varepsilon_i>0$.
 In the case of small disorder, when $U-u-\varepsilon_i\approx U-u$,
this requirement does not essentially restrict our consideration. The conditions of continuity of
the wave function and its first derivative at $x = 0$ allow one to find the constants $A$ and $\varphi$:
 \begin{equation}
 \cases{\psi_U(0)=A\sin\varphi\cr
\psi_U(0)\sqrt{-U}=A\sqrt{U-u}\cos\varphi}
\Rightarrow \hskip5mm
\cases{A^2=|\psi_U(0)|^2{u\over u-U}\cr
\hbox{tg}\varphi=\sqrt{u-U\over U}}
\label{97}
 \end{equation}

The typical values of the wave function module squared in the region of localization $l$, we are interested
in, can be estimated as a half of its peak value within $[0, l]$:
  \begin{equation}
 |\bar\psi_U|^2={A^2\over 2}\hbox{ max}_{x\in[0,l]}\hskip2mm\sin^2[x\sqrt{U-u}+\varphi]
\label{98}
 \end{equation}
 Using Eq.  (\ref{97}), we can obtain for
$ |\bar\psi_U|^2$ the following expressions
  \begin{equation}
 |\bar\psi_U|^2={|\psi_U(0)|^2\over 2}{u\over u-U},
 \hskip10mm\hbox{ when } \hskip10mm l\sqrt{U-u}+\varphi>{\pi\over 2}
\label{100}
 \end{equation}
$$
|\bar\psi_U|^2={|\psi_U(0)|^2\over 2}{u\over u-U}\sin^2[l\sqrt{U-u}+\varphi],
\hskip10mm
\hbox{ when } \hskip10mm l\sqrt{U-u}+\varphi<{\pi\over 2}
$$
 Now we can apply the normalization condition mentioned above $|\bar\psi_U|^2l=1$:
 \begin{equation}
 {|\psi_U(0)|^2 \hskip1mm l\over 2}{u\over u-U}=1,
 \hskip10mm\hbox{ when } \hskip10mm l\sqrt{U-u}+\varphi>{\pi\over 2}
\label{101}
 \end{equation}
\begin{equation}
{|\psi_U(0)|^2 \hskip1mm l\over 2}{u\over u-U}\sin^2[l\sqrt{U-u}+\varphi]=1,
\hskip10mm
\hbox{ when } \hskip10mm l\sqrt{U-u}+\varphi<{\pi\over 2}
\label{102}
\end{equation}
 Equations (\ref{101}), (\ref{102}) and (\ref{95}) allow us to express the localization length $l$ through the
participation function $W(U)$ obtained in this paper. For instance, Eqs. (\ref{101}) and (\ref{95})
  give the following
expression for the localization length:
 \begin{equation}
 l={2|U-u|^{3/2}\over \pi W(U) u^2}\equiv l_0(U)
 \label{103}
 \end{equation}
 For algebraic consistency, we retained here the numerical
factor $2/\pi$. This formula is applicable provided that the localization length $l$,
 obtained with its aid,
meets condition (\ref{101}): $l\sqrt{U-u}+\varphi>{\pi\over 2}$.
 Combining Eqs. (\ref{102}) and (\ref{95}), we obtain equations for determination of the
localization length in the case when $l\sqrt{U-u}+\varphi<{\pi\over 2}$:
 \begin{equation}
 {l_0(U)\over l}=\sin^4\bigg[l\sqrt{U-u}+\hbox{arctg}\sqrt{u-U\over U}\bigg]
 \label{104}
 \end{equation}
 In the topical case when $U-u>\Delta$, formula (\ref{103}) appears
to be the main one. This is why we will not analyze the transcendent equation (\ref{104}).
Figure 2 shows a
typical form of wave functions of the random system at different energies. Horizontal thick lines show
the localization lengths obtained using (\ref{103}) and (\ref{91}).
 It is seen from Fig. 2 that these formulas
may be used to evaluate spectral dependence of the localization length for the states of the random
system considered in this paper.

\section{Conclusions}

The perturbative approach  to  the joint statistics of the advanced and retarded Green's functions,
developed previously for the discrete random 1D models \cite{Koz1,Koz2},
is applied to analysis of the continuous
disordered model described by the Schrodinger equation with a piecewise-constant random potential.
Using the developed approach, we derived the expression for spectral dependence of the degree of
localization in the sense of the Anderson criterion. Numerical verification of the results obtained is
presented. In conformity with the commonly accepted opinion, the states with negative energies of the
considered random system  prove to be, generally speaking, localized, because the participation function
(\ref{91}), at these energies, is nonzero. Exceptions are the points of delocalization arising at
$b\sqrt{-u}>\pi$ (see Eq. (\ref{91})).
 Unfortunately, we have not managed to study behavior of the participation function at large
values of the parameter $b$ (when these points appear), because the used algorithm of numerical solution
of the Schrodinger equation became unstable. In this connection, it makes sense to pay attention to
similarity between the continuous model described in this paper and the discrete model with a complex
structural unit \cite{Koz2}. There are strong grounds for believing that the behavior of the participation
function $W(U)$ of the continuous model under consideration qualitatively coincides with that for the
discrete model \cite{Koz2}, for which the numerical analysis appears to be feasible. In conclusion, emphasize
once again that the developed approach and the results obtained can be useful for analysis of
propagation of the electromagnetic waves in structures of the type of 1D photonic crystals in the
presence of disorder.

\section{Appendix}
\subsection*{ Solution of the spectral problem for the operator ${\cal H}_{\Omega,\eta}$}

To solve this problem, we use the system of eigenfunctions  $\sigma^n_C(x)$ for the operator
 $H_C f(x)\equiv f(C-1/x)/x^2$, obtained in
\cite{Koz3} in explicit form. Consider some of these functions $\sigma(x)$
 and denote the corresponding eigenvalue
of the operator $H_C$ by $\lambda$. Then, the following relationship should be valid
\begin{equation}
{1\over x^2}\sigma(C-1/x)=\lambda\sigma(x)
\label{47}
\end{equation}
 Let us pass, in
this equation, to a new variable $y=[x-A]/B,\hskip3mm x=A+By$,
  where the parameters $A$ and $B$ are supposed to be defined later.
If we now introduce a function $\Phi(y)\equiv C-1/x=C-1/[A+By]$,
 then we can easily see that, by passing to the variable $y$ in
Eq. (\ref{47}), we come to the following relationship
 \begin{equation}
  {1\over B}{d\Phi\over dy}\sigma[\Phi(y)]=\lambda\sigma(A+By)
  \label{51}
  \end{equation}

 Now, we introduce the function $s(y)$ defined as
   \begin{equation}
s(y)\equiv B\hskip1mm\sigma(A+By) \hbox{ and, consequently, }
\sigma(z)={1\over B}\hskip1mm s\bigg({z-A\over B}\bigg)
  \label{52}
  \end{equation}
It follows from (\ref{51}) that
  \begin{equation}
  {1\over B}{d\Phi\over dy}\hskip1mm s\bigg({\Phi(y)-A\over B}\bigg)=\lambda s(y)
  \label{53}
  \end{equation}

 If we define a function ${\cal R}$ by the relation
 \begin{equation}
{\cal R}(y)\equiv {\Phi(y)-A\over B}=
\bigg[{(C-A)A-1\over AB}+{C-A\over A}y\bigg]\bigg/
\bigg[1+{B\over A}y\bigg],
  \label{54}
  \end{equation}
 then Eq. (\ref{53}) can be
rewritten in the form
\begin{equation}
{d{\cal R}\over dy}\hskip1mm s[{\cal R}(y)]=\lambda s(y)
  \label{55}
  \end{equation}

Let us now choose the parameters $A, B$, and $C$ to make operation (\ref{54})
coincident with (\ref{45}). This gives rise to a system of equations for these parameters. Solving this
system we have
\begin{equation}
A={\sqrt{\Omega-\eta}+\imath t\sqrt\Omega\over\sqrt{(\Omega-\eta)(1+t^2)}}
\hskip7mm
B={t\eta\over\sqrt{(\Omega-\eta)(1+t^2)}}
\hskip7mm
C={2\over \sqrt{1+t^2}},\hskip5mm t=\hbox{tg}\bigg[b\sqrt{\Omega-\eta}\bigg]
\label{57}
\end{equation}
 Thus, the function $s(y)$(\ref{52}) constructed with the aid of the eigenfunction $\sigma(x)$ of
the operator $H_C$, for the parameters  $A$, $B$, and $C$  determined by Eq.  (\ref{57})
 is the eigenfunction of the
operator
 ${\cal H}_{\Omega,\eta}$ (\ref{46}),
 with the appropriate eigenvalue $\lambda$ being coincident with that of the operator $H_C$.
Below, we present a compact expression for the $s$-functions.

As shown in \cite{Koz3}, an arbitrary function $f(x)$ may be expanded in series
 in terms of the functions $\sigma^n_C(x)$.
Remind the explicit form of the functions $\sigma^n_C(x)$
and expressions for the eigenvalues  $\lambda_n$  of the operator
$H_C$ at $C < 2$:
 \begin{equation}
 \sigma_C^n(x)={L}_C(x)G^n(x), \hbox{ где }
{L}_C(x)\equiv{1\over 2\pi\imath}\bigg[{1\over x-R}-{1\over x-R^\ast}\bigg],\hskip4mm
 G(x)\equiv\bigg[{R^\ast-x\over R-x}\bigg]
 \label{58}
 \end{equation}
$$
\lambda_n=\bigg({C+\imath\sqrt{4-C^2}\over C-\imath\sqrt{4-C^2}}\bigg)^n\hskip5mm
R\equiv{C+\imath\sqrt{4-C^2}\over 2}\hskip10mm R^\ast\equiv{C-\imath\sqrt{4-C^2}\over 2}\hskip10mm
|C|<2
$$
 as well as the rules of expansion of an arbitrary function $f(x)$ in series in terms of
the above functions:
\begin{equation}
f(x)=\sum_{n=-\infty}^{+\infty}{K}_n\sigma_C^n(x),\hskip10mm {K}_n=\int{f(x)\over G^n(x)}dx
\label{59}
\end{equation}

 Using these relationships, we can obtain similar rules for the eigenfunctions
  (\ref{52})  of the operator(\ref{46})  (called by $s$-functions)
\begin{equation}
{\cal H}_{\Omega,\eta}s_n^{\Omega,\eta}(y)=\lambda_n (\Omega,\eta)s_n^{\Omega,\eta}(y),\hskip5mm
s_n^{\Omega,\eta}(y)=B\hskip1mm \sigma^n_C(A+By),
\hskip5mm\lambda_n(\Omega,\eta)=\exp[2\imath nb\sqrt{\Omega-\eta}]
\label{60}
\end{equation}
 Here, the parameters $A$, $B$, and $C$ are
defined by formulas (\ref{57}).
 The superscript of the $s$-functions indicates their dependence on the energy
argument $\Omega$.

The arbitrary function $f(x)$ can be represented as the series
\begin{equation}
f(y)=\sum_{n=-\infty}^{+\infty}{\cal K}_n s_n(y),\hskip10mm {\cal K}_n=\int{f(y)\over
G^n(A+By)}dy\equiv\int{f(y)\over{\cal G}_{\Omega,\eta}^n(y)}dy,
\label{61}
\end{equation}
with
$$
{\cal G}_{\Omega,\eta}(y)\equiv={r^\ast-y\over r-y},\hbox{ where }
\hskip5mm
r\equiv\hskip1mm{\sqrt\Omega-\sqrt{\Omega-\eta}\over\imath \eta}\hskip5mm
r^\ast\equiv\hskip1mm{\sqrt\Omega+\sqrt{\Omega-\eta}\over\imath \eta}
$$
 Using the quantities
introduced in this way, we may write the following compact expressions for the $s_n^{\Omega,\eta}(y)$
$s$-functions:
\begin{equation}
s_n^{\Omega,\eta}(y)={\cal L}_{\Omega,\eta}(y){\cal G}_{\Omega,\eta}^n(y),\hskip10mm {\cal
L}_{\Omega,\eta}(y)\equiv{1\over 2\pi\imath}
\bigg[{1\over y-r}-{1\over y-r^\ast}\bigg]
\label{62}
\end{equation}
 Here,
the subscripts of the Lorentzian ${\cal L}$  indicate dependence of this function on the energy parameters
$\Omega$ and $\eta=u+\varepsilon$.
 One can easily make sure that
\begin{equation}
{\cal G}_{\Omega,\eta}^n({\cal R}_{\Omega,\eta}(x))=\lambda_n(\Omega,\eta){\cal G}^n_{\Omega,\eta}(x)
\label{621}
\end{equation}



\newpage
\begin{figure}
\begin{center}
\includegraphics [width=10cm]{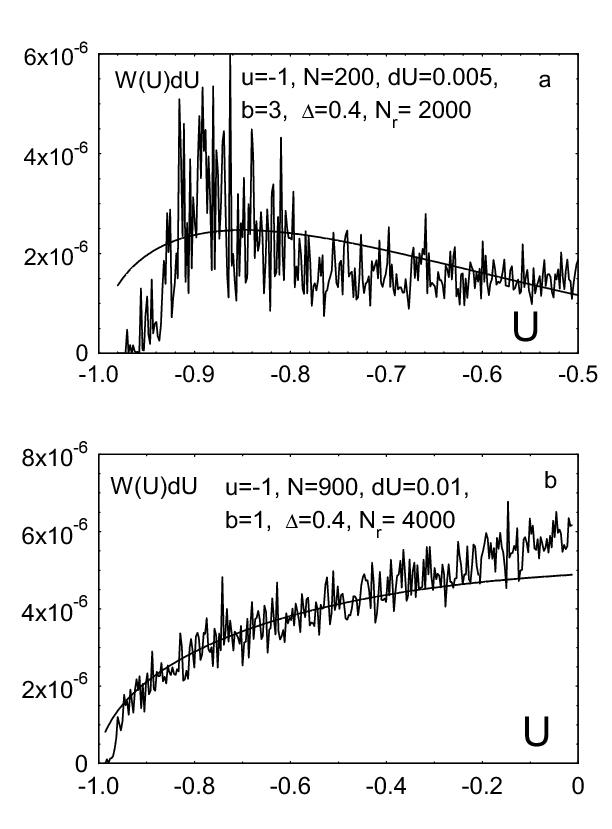}
 \caption{ Spectral dependence of the degree of localization of the states for the 1D disordered system
with a piecewise-constant random potential. The noisy curves are
obtained by computer simulation and the smooth ones are computed
using Eq.  (\ref{91}).}
 \label{fig1}
 \end{center}
\end{figure}

\begin{figure}
\begin{center}
\includegraphics [width=10cm]{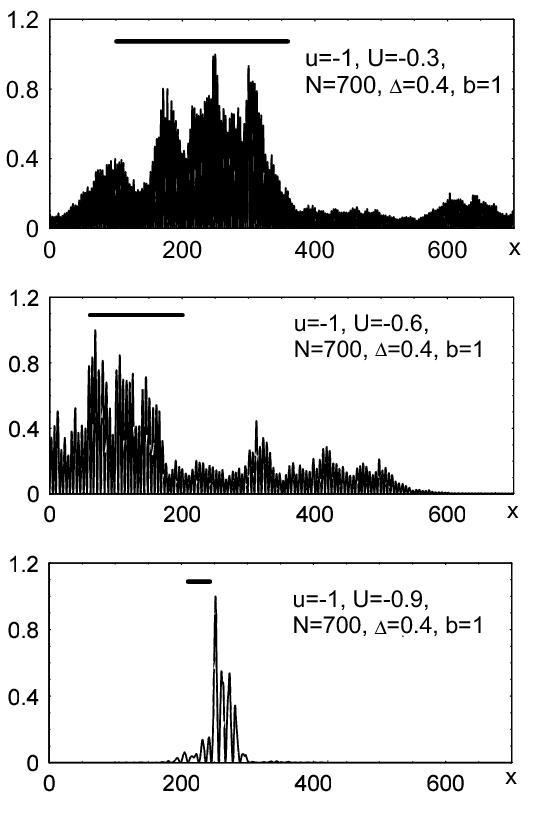}
 \caption{ Wave function of the 1D disordered system with a piecewise-constant random potential for different
 values of the energy $U$. Thick horizontal lines show the
 localization length calculated using Eq. (\ref{103}).}
 \label{fig2}
 \end{center}
\end{figure}



\begin{thebibliography}{99}
\bibitem{LGP}  I.M.Lifhits, S.A.Gredeskul, and L.A.Pastur, {\it
Introduction to the Theory of Disordered Systems} [in Russian] Nauka, Moscow (1982);
English transl., Wiley, New York (1988).
\bibitem{Belous1}   M.V.Belousov and D.E.Pogarev, {\it JETP Lett.,} {\bf 36}, 189 -191(1982).
\bibitem{Belous2}  M.V.Belousov, B.E.Vol'f, and E.A.Ivanova,
{\it JETP Lett.,} {\bf 38}, 456 -- 459 (1983).
\bibitem{Dyson}  Dyson F.J. Phys.Rev., {\bf 92}, p. 1331, (1953).
\bibitem{Ber1}  V.L.Berezinsky, {\it JETP}, {\bf 65}, p. 1251 (1973).
\bibitem{Koz1} G.G.Kozlov {\it Theoretical and Mathematical Physics}, {\bf 162}(2): 238 -- 253 (2010).
\bibitem{Koz2} G.G.Kozlov, accepted to {\it Applied Mathematics}, (2011).
\bibitem{And}  Anderson P.W., Phys.Rev., {\bf 109}, p. 1492, (1958).
\bibitem{Koz3}  Kozlov G.G.,
"Spectrum and eigen functions of the operator $H_Uf(x)\equiv f(U-1/x)/x^2$ and strange attractor's
density for the mapping $x_{n+1}=1/(U-x_n),$" arXiv:0803.1920. [math-ph] (2008).
\end{thebibliography}
\end{document}